# Coexistence of Ferromagnetism and Topology by Charge Carrier Engineering in intrinsic magnetic topological insulator MnBi$_4$Te$_7$

Bo Chen[1,2,†], Fucong Fei[1,2,†,*], Dinghui Wang[1,†], Zhicheng Jiang[3,†], Bo Zhang[4], Jingwen Guo[1,2], Hangkai Xie[1,2], Yong Zhang[1,2], Muhammad Naveed[1,2], Yu Du[1,2], Zhe Sun[4], Haijun Zhang[1,*], Dawei Shen[3,*], Fengqi Song[1,2,*]

[1]National Laboratory of Solid State Microstructures, Collaborative Innovation Center of Advanced Microstructures, and College of Physics, Nanjing University, Nanjing 210093, China.

[2]Atomic Manufacture Institute (AMI), Nanjing 211805, China

[3]Center for Excellence in Superconducting Electronics, State Key Laboratory of Functional Materials for Informatics, Shanghai Institute of Microsystem and Information Technology, Chinese Academy of Sciences, Shanghai 200050, China

[4]National Synchrotron Radiation Laboratory, University of Science and Technology of China, 230029 Hefei, China

†These authors contributed equally to this work. *Correspondence and requests for materials should be addressed to F. F (email: feifucong@nju.edu.cn); H. Z. (email: zhanghj@nju.edu.cn); D. S. (dwshen@mail.sim.ac.cn); or F. S. (email: songfengqi@nju.edu.cn).

Abstract

Intrinsic magnetic topological insulators (MTIs) MnBi$_2$Te$_4$ and MnBi$_2$Te$_4$/(Bi$_2$Te$_3$)$_n$ are expected to realize the high-temperature quantum anomalous Hall effect (QAHE) and dissipationless electrical transport. Extensive efforts have been made on this field but there is still lack of ideal MTI candidate with magnetic ordering of ferromagnetic (FM) ground state. Here, we demonstrate a MTI sample of Mn(Bi$_{0.7}$Sb$_{0.3}$)$_4$Te$_7$ which holds the coexistence of FM ground state and topological non-triviality. The dramatic modulation of the magnetism is induced by a charge carrier engineering process by the way of Sb substitution in MnBi$_4$Te$_7$ matrix with AFM ordering. The evolution of magnetism in Mn(Bi$_{1-x}$Sb$_x$)$_4$Te$_7$ is systematically investigated by magnetic measurements and theoretical calculations. The clear topological surface states of the FM sample of $x = 0.3$ are also verified by angle-resolved photoemission spectra. We also aware that the FM sample of $x = 0.3$ is close to the charge neutral point. Therefore, the demonstration of intrinsic FM-MTI of Mn(Bi$_{0.7}$Sb$_{0.3}$)$_4$Te$_7$ in this work sheds light to the further studies of QAHE realization and optimizations.

**Introduction**

Topological insulators, combined with the magnetism, can host various intriguing phenomena, such as quantum anomalous Hall effect (QAHE), topological electromagnetic effect, axion state and so on[1-6]. However, due to the confinement of the rare materials candidates, the research of the magnetic topological insulators (MTIs) is limited in the magnetic doped topological insulators or the magnetic proximity in heterostructures[7-10]. In recent years, intrinsic magnetic topological insulators $MnBi_2Te_4$ and a series of $MnBi_2Te_4/(Bi_2Te_3)_n$ materials have sprung up, in which the spontaneous magnetization derives from the internal manganese atoms instead of the external impurities[11-17]. Unfortunately, because of the compensated antiferromagnetic (AFM) ground state, the intensity of magnetism in $MnBi_2Te_4$ is rather weak under the zero field. $MnBi_2Te_4/(Bi_2Te_3)_n$ series can significantly reduce the interlayer AFM exchange interaction by intercalated $Bi_2Te_3$ quintuple layers into $MnBi_2Te_4$ martrix[18,19], but the AFM ground state still possesses the lower energy than ferromagnetic (FM) state and a small magnetic field is still essential to align the opposite magnetic moments. Furthermore, in all of these recent known intrinsic MTI candidates, the ineluctable antisite defects between the manganese and bismuth atoms will induce the heavy n-type trivial bulk carriers[13,20], covering the transport of the novel chiral edge state. Up to now, QAHE can only be realized at $MnBi_2Te_4$ thin films with the high magnetic field up to 6 T from three to eight septuple layers (SLs) or near zero field with the high bias voltage (~200V) at 5 SLs[21-23]. Ultra-high magnetic field due to AFM ground state and the huge bias voltage resulting from the antisite defects obstruct the further application of

MnBi$_2$Te$_4$ in the AFM spintronics and dissipationless quantum devices. In MnBi$_2$Te$_4$, the Sb substitution at the Bi sites can neutralize the excess bulk carrier efficiently, while the AFM state is still robust[24,25]. However, in the case of MnBi$_4$Te$_7$, the comparable AFM-FM energy difference paves the way to modify the AFM state and carrier density at the meantime, with no-magnetic elements doping.

In this work, we grow a serials single crystals of Mn(Bi$_{1-x}$Sb$_x$)$_4$Te$_7$ from $x = 0.15$ to $x = 0.46$. Angle-resolved photoemission spectra (ARPES), magnetic and electrical transport measurement are conducted to explore the influence of the Sb substitution in Mn(Bi$_{1-x}$Sb$_x$)$_4$Te$_7$. Via no-magnetic element Sb doping, the AFM ground state evolves into ferromagnetic (FM) state at $x = 0.3$ and return to AFM when continuously increasing Sb ratio to $x = 0.46$, hosting an unexpected FM region in the middle. According to the theoretical calculation, the energy difference between AFM and FM in Mn(Bi$_{1-x}$Sb$_x$)$_4$Te$_7$ family is only ~ 0.2 meV, indicating that the magnetic ordering could be easily affected by external conditions. The calculation also reveals a distinct dip of the energy difference value between the AFM and FM at $x \sim 0.3$, qualitatively matching with the trend displayed in our experimental results. Furthermore, according to the electrical transport, we find that similar to the case of MnBi$_2$Te$_4$[24,25], Sb doping is also an effective way to modulate the carrier densities. Specifically, the heavy n-type bulk carriers in pure MnBi$_4$Te$_7$ are significantly suppressed by Sb doping, and an n-p transition occurs between $x = 0.36$ and 0.4. One may notice that the electrical neutral point of Sb doping MnBi$_4$Te$_7$ is close to the FM state region. In ARPES measurement, the clear surface Dirac cone of the Bi$_2$Te$_3$ termination prove the robust topological property at $x = 0.3$ in Mn(Bi$_{1-x}$Sb$_x$)$_4$Te$_7$. Our work demonstrates the obvious engineering of the both magnetism and the charge carriers in MnBi$_4$Te$_7$ by Sb doping. Mn(Bi$_{1-x}$Sb$_x$)$_4$Te$_7$ with $x$ around 0.3 is a promised candidate of MTI with FM ground state and low bulk carrier concentrations, which provides a prospective avenue to accomplish the QAHE at the zero field as well as other exotic topological effects..

**Results**

Mn(Bi$_{1-x}$Sb$_x$)$_4$Te$_7$ is a layered rhombohedral material with the space group P$\bar{3}$m1 (164)[26-29]. The Mn(Bi$_{1-x}$Sb$_x$)$_2$Te$_4$/(Bi$_{1-x}$Sb$_x$)$_2$Te$_3$ superlattice stacks along the c axis through the weak van der Waals forces as is drawn in Figure 1(a). There are two various cleavage planes with independent band structures and Dirac cones as reported before: the Mn(Bi$_{1-x}$Sb$_x$)$_2$Te$_4$ SL and the (Bi$_{1-x}$Sb$_x$)$_2$Te$_3$ quintuple layer[19,28,30-32]. We can synthesize the crystals of Mn(Bi$_{1-x}$Sb$_x$)$_4$Te$_7$ by self-flux method to obtain the lustrous flakes with the average size about 3 × 3 mm$^2$. The inset of the Figure 1(b) is the morphology image of the as-grown shiny Mn(Bi$_{1-x}$Sb$_x$)$_4$Te$_7$ crystals. Single crystal X-ray diffraction (XRD) patterns show prominent peaks labelling (00n) Miller indexes in Figure 1(b). The refined value of the c axis is nearly at ~ 24 Å with the tiny change at different doping ratio, indicating the similar structure constants and identical layered-crystal structure of the all samples. Figure 1(c) displays the energy dispersive spectra (EDS) which shows the stoichiometric ratio of Mn: (Bi + Sb): Te is about 1: 4: 7. The relative intensity of Bi at 3.8 keV and Sb at 3.6 keV evolve as expected after normalizing the counts with the Te peaks, elucidating the valid substitution of the two elements. Both the XRD and EDS results demonstrate the favourable crystallinity and the precise molar ratio during the procedure of the substitution in Mn(Bi$_{1-x}$Sb$_x$)$_4$Te$_7$.

To explore the magnetic variation in the Mn(Bi$_{1-x}$Sb$_x$)$_4$Te$_7$ samples, we conduct the magnetic measurement of the field cool (FC) and zero-field cool (ZFC) processes from 300 K to 2 K by vibrating sample magnetometer. The magnetic field is applied perpendicular to the sample cleave surfaces with the magnitude at 500 Oe. In the pure MnBi$_4$Te$_7$, the FC-ZFC curves show the expected antiferromagnetic peaks at 13 K and the overlapped ones diverge at lower temperature (~ 5K), corresponding with previous articles[31,33]. The magnetic field dependence of the magnetization curves exhibit the zigzag shape at 2 K and the hysteresis loop divides into two jump points deviating from the zero field at 10 K. Compared with the MnBi$_2$Te$_4$ (~ 7T)[13,20], the saturation field to approach the FM region in the MnBi$_4$Te$_7$, less than 0.3 tesla, is much smaller. But as is shown in the Figure 2(b), in the as-grown MnBi$_4$Te$_7$, the magnetic ground state is still AFM and external magnetic field is still essential to form FM state.

Due to the Bi$_2$Te$_3$ intercalation which enlarges the distance between the neighboring MnBi$_2$Te$_4$ SL, the energy difference between the AFM and FM is relatively small in MnBi$_4$Te$_7$[18,19,34]. Weak interlayer AFM coupling paves the way for the magnetism modulation by various approaches. For example, in the case here, when substituting Bi with the no-magnetic Sb element, the AFM ground state can be altered availably without introducing the undesired outer magnetic impurities. From the FC-ZFC curves of the doping samples in Figure 2(a), the magnitudes of the divergence at lower temperature (~ 5K) rise and excess the value of the AFM transition point at 13 K, proving the observably enhancement of the FM ingredient compared with $x = 0$. Most importantly, the AFM peaks is absent at the specific ratio of the $x = 0.3$ and FC-ZFC curves diverge directly at 13 K, indicating the signature of the standard FM state. Along with $x$ increasing continuously, the FM at 13 K disappears at $x = 0.36$ and 0.4, showing the recovery of AFM while at $x = 0.46$, the FC-ZFC curves display the peculiar ferrimagnetic character.

    To further explore the property of the AFM and FM states of the various doping ratios, magnetization versus magnetic field (M-H) measurement is conducted shown in Figure 2(b). The sizes of the hysteresis loop and the corresponding values of the coercive field decrease monotonously with the increase of the Sb concentration. However, the FM sample of $x = 0.3$ is an exception and holds a significant larger coercive field than $x = 0.15$. Further detailed discussion on this sample will be carried out below. After doping Sb element, the feature of zigzag shaped hysteresis loops at 2 K and zero field evolves to the inclined rectangle shaped loops. However, at 10 K, the hysteresis loop still evolves into two parts, indicating the recovered domination of the AFM feature at high temperature ($x = 0.3$ is still an exception on this point). In the $x = 0.3$ sample, the hysteresis at 2 K is similar to the ones in the samples of $x = 0.15$ and $x = 0.36$. However, when increasing the temperature, the loop does not split into two loops with zigzag shaped M-H curve but simply shrinks and is always pinned at the zero field. This kind of hysteresis behavior of demonstrates a typical FM ordering hold in $x = 0.3$ sample, consistent with the FC-ZFC behavior discussed above. In addition, due to the distinct magnetic ordering, the coercive field in this sample is anomalously large comparing

with other samples (see Figure 5(a)). When the $x$ ratio extends into 0.36 and 0.4, the AFM state recurs and the AFM transition points emerge. In the ratio of 0.46, the M-H curves display the possible ferrimagnetic characters. Since the magnetism of the samples shows no-monotonous and complex tendency with the Sb doping, multifold internal mechanisms may be responsible for the complex evolution of the magnetism.

We feel excited to accidentally discover the sample holding FM ground state when finely tuning the Sb doping ratio around $x = 0.3$, but the Sb substitution undoubtedly weakens the spin-orbital coupling comparing with pure $MnBi_4Te_7$, and a topological transition may occur. Thus further identification of the topological nature of this FM sample with $x = 0.3$ is essential. In Figure 3(a)-(d), we display the ARPES measurement results of $Mn(Bi_{0.7}Sb_{0.3})_4Te_7$ at the photon energies of 6.6 eV, 7.25 eV, 9 eV and 20 eV. It is obvious to distinguish a linear Dirac surface state at the termination of $Bi_2Te_3$ quintuple layers which is independent of the photon energies. Therefore, the no-trivial topological property is still retentive at $x = 0.3$. The surface Dirac cone show the no-gap feature which may be attributed to the modified hybridization between topological surface state and the valence bands after Sb doping[28,35]. To obtain the exact position of the Dirac point, we map the constant energy contours in Figure 3(e). Compared with the pure $MnBi_4Te_7$ (~ 0.3eV), Dirac point is near the Fermi level at $x = 0.3$, about 0.2 eV. The tendency of Fermi level descending indicates the relative low carrier density in $x = 0.3$. It is apparent that the no-trivial topological surface state is still held, confirming the topological property in the FM region at $x = 0.3$.

Accompanying by the engineering of the magnetism, Sb doping can also modify the carrier density of MnBi$_4$Te$_7$. Due to the internal defects of Mn and Bi antisite, the as-grown MnBi$_2$Te$_4$ and related heterostructure MnBi$_4$Te$_7$ are serious n-type doped regardless of the growth methods[13,20]. Heavy bulk carriers will obstruct the regulating capacity of the gate voltage which confines the device experiments at ultra-thin films. According to the previous efforts in topological insulators[25,36], no-magnetic element doping is an efficient way to adjust the Fermi level of the samples without destructing the topological property and the carrier mobility.

The magneto resistivity and Hall resistivity of the Mn(Bi$_{1-x}$Sb$_x$)$_4$Te$_7$ are shown in Figure 4. Resistivity versus temperature curves in Figure 4(a) display the characteristic kinks at near 13 K, corresponding to the Néel and Curie temperature obtained from the magnetic measurement. In Figure 4(b), the carrier densities are extracted from the Hall resistivity at 20 K. We can find that the n-p transition occurs at the ratio between $x = 0.36$ and 0.4 in which the symbol of the Hall coefficient changes from negative to positive. The carrier densities in $x = 0.36$ and 0.4 are $1.65 \times 10^{19}$ cm$^{-3}$ (n type) and $2.6 \times 10^{19}$ cm$^{-3}$ (p type) respectively, ten times smaller than the value in pure MnBi$_4$Te$_7$. Since the electrical neutral point has been verified, we then take more attention to the magneto transport behavior of the samples which are near the electrical neutral point. As is shown Figure 4(c) and (d), the longitudinal and transverse resistivity show the clear butterfly curves and hysteresis loops due to the spontaneous magnetization. Because of the complex magnetic competition between the AFM and FM state, the resistivity versus magnetic field curves display the complex evolution trend near the neutral point from $x = 0.3$ to $x = 0.4$. It is worth noting that the Hall signal changes the sign from $x = 0.3$ to $x = 0.46$. Similar results has been discovered in MnBi$_2$Te$_4$ and is believed to be attributed to the competition of intrinsic Berry curvature and extrinsic skew scattering[37]. The AHE of $x = 0.36$ displays a queer behavior with round shaped loops, probably because it is close to the transition point from negative to positive AHE. Similar with the M-H curves, the hysteresis loop of the Hall resistivity separates into two individual loops at the 10 K except the sample of $x = 0.3$ Sb doping. In the sample of $x = 0.3$ with FM state, the AHE is still observable near the zero field at 10 K corresponding with the hysteresis loop in M-H signals. The negative magnetoresistivity even above the Néel and Curie temperature could result from the strong spin fluctuation[13,38]. Contrary to stubborn AFM in the Mn(Bi$_{1-x}$Sb$_x$)$_2$Te$_4$[24,25], it is apparent that the FM ground state can be approached accompanied with the successful modulation of carrier densities in Sb-doped MnBi$_4$Te$_7$ samples.

To show and elucidate the properties of the magnetism and the electrical transport during the procedure of doping, evolution diagrams are extracted and drawn from the curves of M-H and Hall coefficients. The anomalous Hall ratio ($\Delta\rho_{xy}$-0T)/($\Delta\rho_{xy}$-saturation) as well as the coercive field of the six samples, are displayed in Figure 5(a). Compared with the pure MnBi$_4$Te$_7$ sample (~ 0.066 T), the sizes of coercive fields decrease monotonously after doping except an obvious peak at $x = 0.3$. However, in the contrary, the AHE ratio enhances dramatically under antimony substitution. In the pure MnBi$_4$Te$_7$, the ratios are nearly about 25% at 2 K and 0% at 5 K respectively. In $x = 0.15$ and 0.3 under 2 K, the ratios rise into 99%, implying the ideal saturated magnetic polarization under zero field. Most importantly, upon increasing the temperature, the

AHE ratio at $x = 0.3$ under zero field can stabilize at about 75%, excessing the sizes of $x = 0.15$ and 0.36, about 30%, not to mention the near zero value in pure MnBi$_4$Te$_7$. It is obvious that the AFM transform into FM at $x = 0.3$, with the abnormal high coercive field and AHE ratio, ensuring the strong spontaneous magnetization.

Hereinbefore, we conduct the discussion on the experimental results of the magnetism and carriers, especially the FM state at $x = 0.3$. To explore and understand the internal mechanism of the magnetism transition qualitatively in theory, we conduct the first-principle calculations from $x = 0$ to 0.5. The energy difference curve of AFM and FM shows a clear dip with the minority point at $x = 0.3$, corresponding with the site of the FM region as displayed in Figure 5(b). However, the value of the energy difference is still about 0.2 meV, far away from the balanced point between AFM and FM. Besides, at higher doping ratio $x = 0.46$, the FC-ZFC curves display a co-existence phase of the FM and ferrimagnetism, inconsistent with expected growing AFM state in Figure 5(b). Therefore, another factor may play a no-negligible role and involve in the magnetic competition, which deserves further consideration. In the pure MnBi$_2$Te$_4$ and MnBi$_4$Te$_7$, antisite defects between Mn and Bi atoms are believed to exist[13,20,39]. Although we substitute the Bi with Sb exactly, a few Sb atoms will inevitably participate in the antisite replacement with Mn atoms. According to the theoretical prediction, Mn/Sb antisite will decrease the energy difference between FM and AFM markedly[40,41]. In the Mn(Bi$_{1-x}$Sb$_x$)$_4$Te$_7$ samples, antisite defects between Sb and Mn atoms will shift down the energy difference curve in the whole, which will create a narrow zone of negative value, account for the FM region at $x = 0.3$. Along with the further increase of the doping ratio, the influence of the Mn/Sb defects will enhance, which is the origin of the emergence of the ferrimagnetism at $x = 0.46$[42]. It is obvious that the internal mechanisms are complex in Mn(Bi$_{1-x}$Sb$_x$)$_4$Te$_7$. Due to the close energy difference between the AFM and FM, the magnetic ground state can be affected and disturbed by the tiny perturbations. Although the theoretical calculations cannot exactly match with the experiment results, the overall tendency extracted from the first principle calculation can be useful to investigate and understand the anomalous magnetic transition at $x \sim 0.3$.

**Conclusion**

In summary, the magnetism and the carrier density have been modulated simultaneously at the Mn(Bi$_{1-x}$Sb$_x$)$_4$Te$_7$ crystal. The AFM ground have evolved into FM at $x = 0.3$, returned to AFM at $x = 0.36$ and then turned into ferrimagnetic at $x = 0.46$. Multiple interactions including spin-orbit coupling, FM-AFM coupling and Mn/Sb antisite involve and contribute the general magnetism. Besides, the dominated carriers can be modified from n-type at $x = 0.36$ to p-type $x = 0.4$, to neutralize the undesirable bulk carriers near the FM region. At the region from $x = 0.3$ to 0.4, there are intriguing magnetic and charge carriers transitions, providing an excellent avenue to study the interaction coupling between the magnetism and carriers. Mn(Bi$_{1-x}$Sb$_x$)$_4$Te$_7$ with $x$ around 0.3 is a promised candidate of MTI with FM ground state and low bulk carrier concentrations, and this material displays improved performance on both magnetism and electrical transport than pure MnBi$_4$Te$_7$. Our results shed light to the exploration of intrinsic MTI candidates with expected performance and the related studies such as QAHE realization and optimizations.

*Note Added*: During preparing this manuscript, we notice another independent work (arXiv: 2008.09097) also focusing on Mn(Bi$_{1-x}$Sb$_x$)$_4$Te$_7$ but with wider Sb doping ratio range[42]. These two works complement each other, and combing them can get more comprehensive understanding on this material system.

**Acknowledgments:**

The authors gratefully acknowledge the financial support of the National Key R&D Program of China (2017YFA0303203, 2016YFA0300204), the National Natural Science Foundation of China (Grant Nos. 91622115, 11522432, 11574217, U1732273, U1732159, 61822403, 11874203, 11904165, 11904166, 11674165 and 11227902), the Natural Science Foundation of Jiangsu Province (BK20190286), Users with Excellence Program of Hefei Science Center CAS (2019HSC-UE007), the Fundamental Research Funds for the Central Universities (020414380082, 020414380150, 020414380151, 020414380152 and 020414380038), the opening Project of the Wuhan National High Magnetic Field Center, the Fok Ying-Tong Education Foundation of China (161006), High Performance Computing Center of Collaborative Innovation Center of Advanced Microstructures and Award for OutstandingMember in Youth Innovation Promotion Association CAS.

**Figure Captions**

**Fig.1 Characterization of the Mn(Bi$_{1-x}$Sb$_x$)$_4$Te$_7$.** (a) Crystal structure of the Mn(Bi$_{1-x}$Sb$_x$)$_2$Te$_4$/(Bi$_{1-x}$Sb$_x$)$_2$Te$_3$ superlattice. (b) Single crystal X-ray diffraction of the four samples. Inset: Optical images of the as-grown sample. (c) Energy dispersive spectra after normalized with the intensity of the Te element.

**Fig. 2 Magnetism measurement of the Mn(Bi$_{1-x}$Sb$_x$)$_4$Te$_7$.** (a) FC-ZFC curves of the six samples. The curves are offset for clarity. (b) The magnetic field dependence of the magnetization from 2 K to 20 K.

**Fig. 3 ARPES images of the Mn(Bi$_{1-x}$Sb$_x$)$_4$Te$_7$ at $x$ = 0.3.** (a)-(d) ARPES measurement at the photon energies of 6.6 eV, 7.2 eV, 9 eV and 20 eV respectively. The topological surface states are marked by the red arrows. (e) Constant-energy maps at binding energies from -0.02 to -0.22 eV with the photon energy of 6.6 eV.

**Fig. 4 Electrical transport of the Mn(Bi$_{1-x}$Sb$_x$)$_4$Te$_7$.** (a) Resistivity versus temperature from $x$ = 0.15 to 0.46. (b) Carrier density extracted from the Hall signals at 20 K. (c) Magneto resistivity and (d) the anomalous Hall after subtracting the background at the selected ratios: $x$ = 0.3, 0.36 and 0.4.

**Fig. 5 Evolution diagrams of the magnetism and AHE.** (a) AHE ratio at 2 K and 5 K

and the coercive field of the samples. (b) FM-AFM energy difference upon Sb doping. The blue stars represent the FM region at the ratio of $x = 0.3$ Sb concentration.

**Figures**

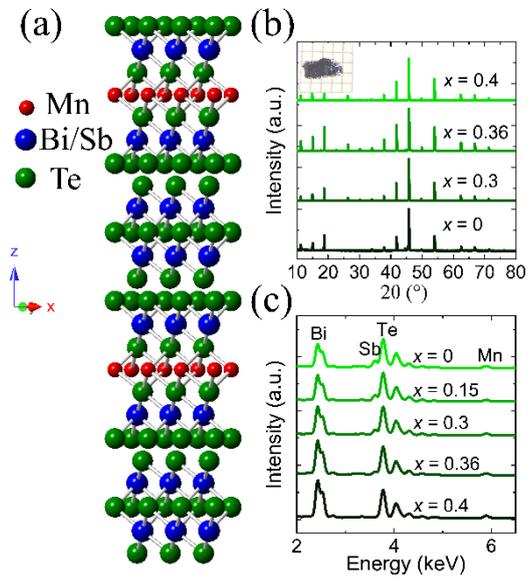

**Fig.1 Characterization of the Mn(Bi$_{1-x}$Sb$_x$)$_4$Te$_7$.**

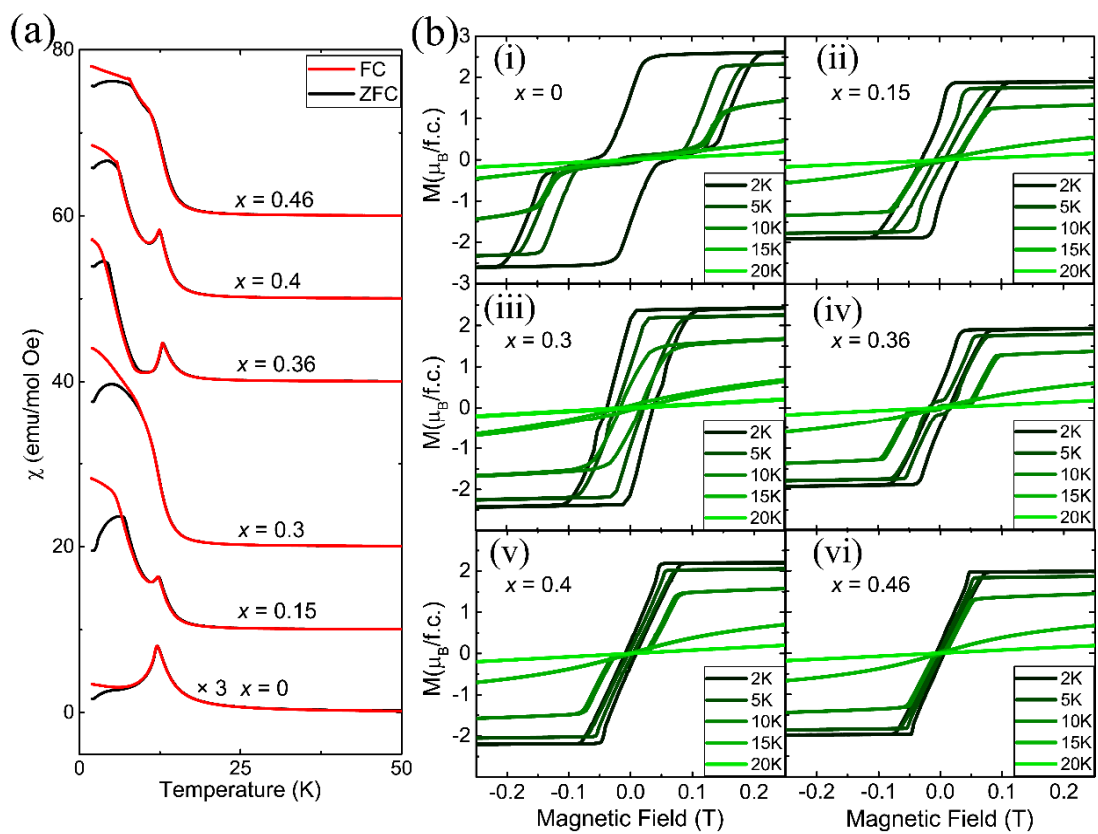

**Fig. 2 Magnetism measurement of the Mn(Bi$_{1-x}$Sb$_x$)$_4$Te$_7$.**

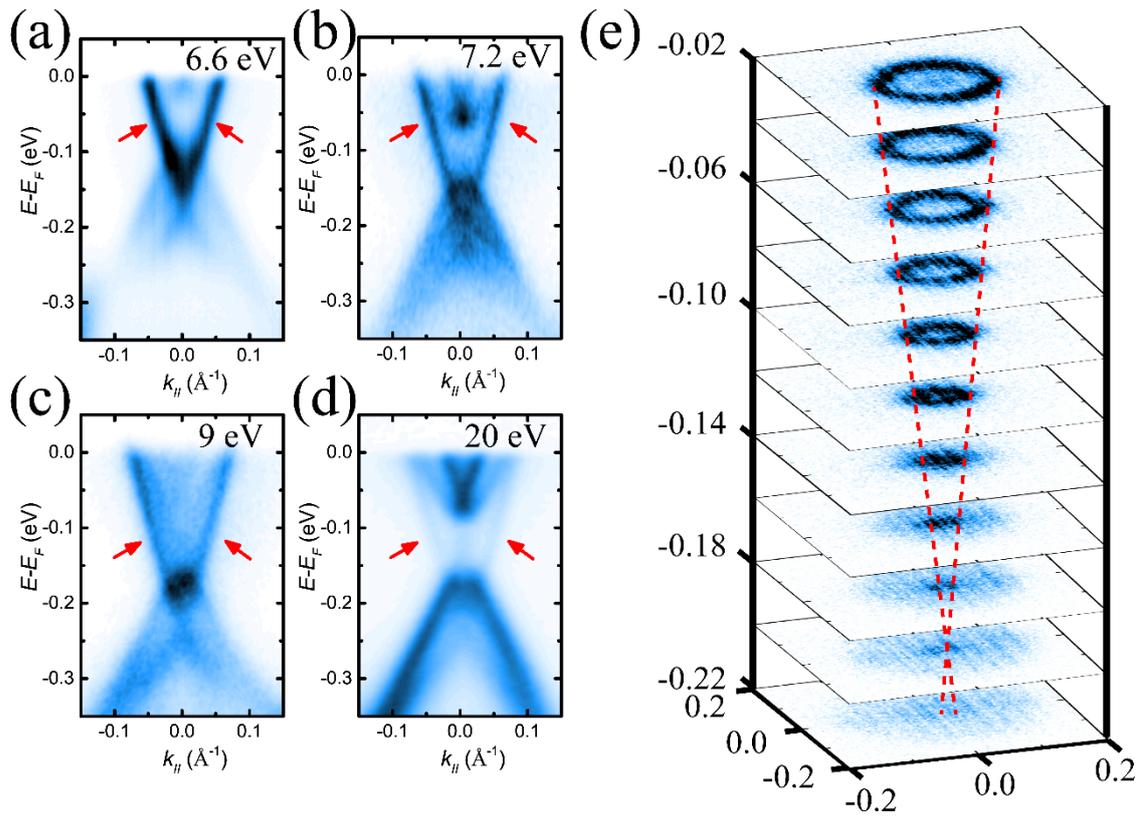

**Fig. 3** ARPES images of the Mn(Bi$_{1-x}$Sb$_x$)$_4$Te$_7$ at *x* = 0.3.

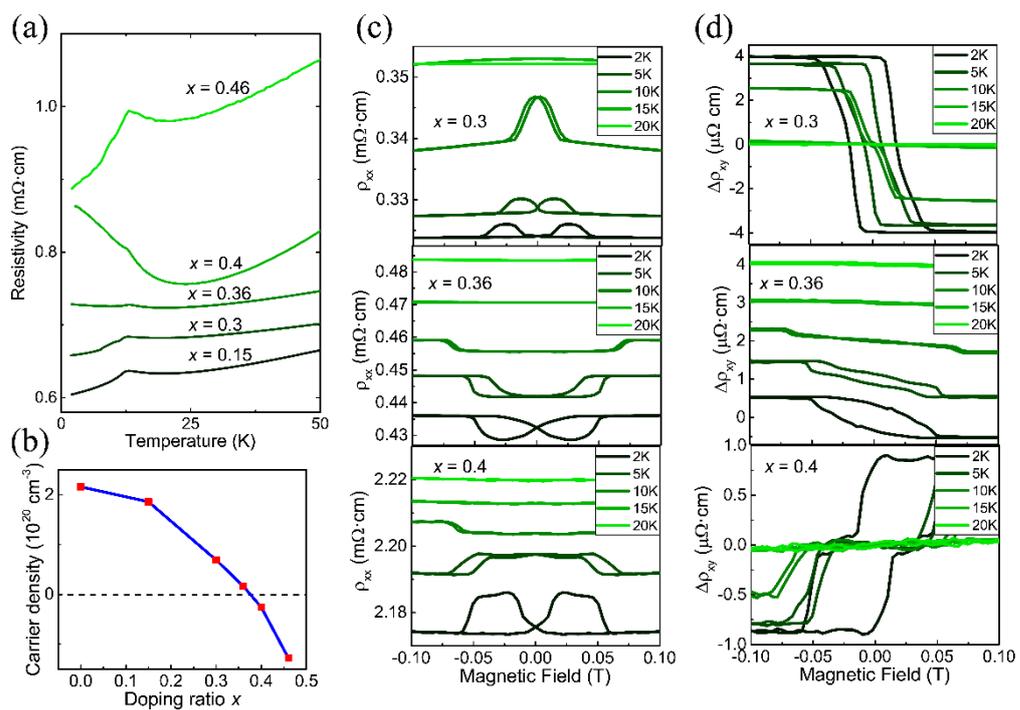

**Fig. 4 Electrical transport of the Mn(Bi$_{1-x}$Sb$_x$)$_4$Te$_7$.**

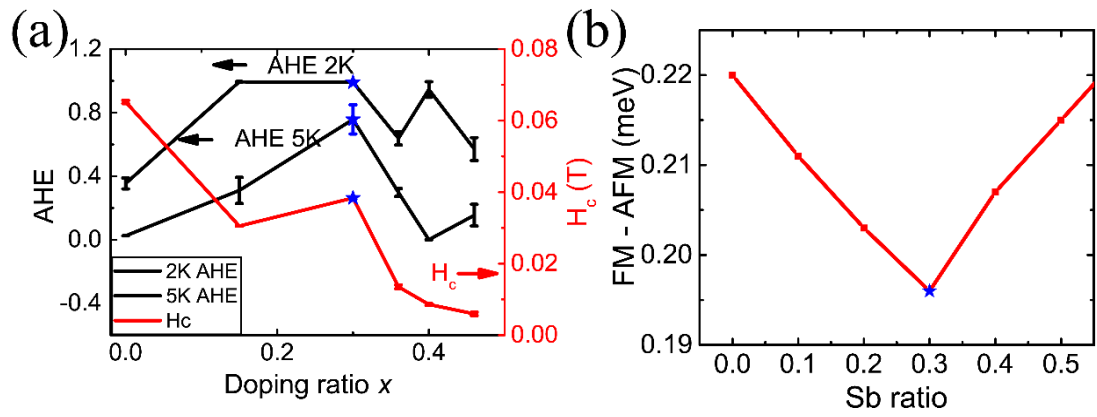

**Fig. 5 Evolution diagrams of the magnetism and AHE.**